\documentclass[12pt]{iopart}

\usepackage{bm}
\usepackage{array}
\usepackage{graphicx}
\usepackage{dcolumn}  
\usepackage{amssymb}
\usepackage[english]{babel}

\begin{document}

\title{The passive diffusion of \textit{Leptospira interrogans}}
\date{\today}
\author{Lyndon Koens, and Eric Lauga\footnote{e.lauga@damtp.cam.ac.uk}}
\address{ Department of Applied Mathematics and Theoretical Physics, University of Cambridge, Wilberforce Road, Cambridge CB3 0WA, United Kingdom}
\begin{abstract} 
Motivated by recent experimental measurements, the passive diffusion of the bacterium  \textit{Leptospira interrogans} is investigated theoretically. By approximating the cell shape as a straight helix and using the slender-body-theory approximation of Stokesian  hydrodynamics, the resistance matrix of \textit{Leptospira} is first determined numerically. The  passive diffusion of the helical cell is then obtained computationally using a  Langevin formulation which is sampled in time in a manner consistent with the experimental procedure. Our results are in excellent quantitative agreement  with   the  experimental results with no adjustable parameters. 
\end{abstract}
\maketitle

\section{Introduction}

Small particles in a fluid undergo a continuous random displacement called Brownian motion. This phenomenon was first studied experimentally in the 19th century by Brown \cite{brown1828xxvii}, at the origin of its name,  and first explained theoretically in 1905 by Einstein \cite{Einstein1905}. Einstein did this by considering the equilibrium behaviour of a suspension of spheres under an unidirectional force. He compared the equation from the minimization of the thermodynamic free energy to the advection-diffusion equation for the concentration of the spheres. The common form of these equations enabled him to determine the diffusion coefficient of a sphere in one dimension. Shortly thereafter, Langevin showed that the same result could be obtained using Newton's second law if an appropriate fluctuating force, $\mathbf{F}_{Br}$, is applied to the body \cite{Langevin1908}. In a typical micron-size system, and if very short time scales are not of interest,  inertial forces can be neglected   and  Langevin's equations for a rigid body simplify to the so-called Brownian Dynamics approach as
\begin{eqnarray}
\mathbf{\tilde{F}}_{H} +  \mathbf{\tilde{F}}_{Br} &=&\mathbf{0}, \label{forcebal} \\
\left\langle\mathbf{\tilde{F}}_{Br}\right\rangle &=& {\bf 0}, \label{Fmean} \\
\left\langle\mathbf{\tilde{F}}_{Br}(0) \otimes  \mathbf{\tilde{F}}_{Br}(t)\right\rangle &=& 2 k_{b} T \mathbf{R}_{\tilde{F} \tilde{U}} \delta(t), \label{Fcovars}
\end{eqnarray} 
where $\mathbf{\tilde{F}}_{H}$ is a six-component vector containing the instantaneous hydrodynamic forces and torques on the body, $\mathbf{\tilde{F}}_{Br}$ denotes the forces and torques created by thermal fluctuations, $\left\langle\cdot \right\rangle$ represents   ensemble averaging,  and $\mathbf{R}_{\tilde{F} \tilde{U}}$ is the full (6-by-6) hydrodynamic resistance matrix of the rigid body of interest.  Notation-wise, in the equations above,  $\otimes$ denotes the outer tensor product, $t$ is time, $k_{b}$ is the Boltzmann constant, $T$ is the absolute temperature,  and $\delta(t)$ is the Dirac delta function. Einstein's and Langevin's results have since been extended to look at the diffusion  of rods \cite{Prager1955}, helicoidal bodies \cite{Brenner1965}, helices \cite{Hoshikawa1979} and arbitrarily shapes \cite{Brenner1967,Ermak1978,Makino2004,Sun2008}. The influence of fluid and particle inertia was also theoretically investigated \cite{Hinch1975}, predicting the  short time  Brownian ballistic regime. Only recently, improvements in imaging techniques have allowed for experimental investigations of the Brownian ballistic regime \cite{Lukic2005,Huang2011,Li2013} and the diffusion of shapes other than spheres \cite{Han2006,Mukhija2007,Butenko2012,Kraft2013}. 

Recent work in the biophysics of swimming microorganisms has  raised interest into the diffusion of active particles which  use internal processes to swim through the fluid. Many bacteria are classic examples of active particles. The diffusive behaviour of such bodies has recently been explored experimentally \cite{Kummel2013,Buttinoni2013} and theoretically  \cite{Eri2011,TenHagen2011,Wittkowski2012,Mijalkov2013}. The active motion (typically self-propelled swimming)  increases the effective diffusion of the particles by orders of magnitude, even if said swimming motion would produce no net displacement in the absence of thermal fluctuations \cite{Eri2011}. These theoretical investigations typically simplify the shape of the active particle  to that of a sphere or an ellipsoid. However,  it is the dynamics of these more complicated shapes which typically result in the motion of the  active particle. For example \textit{Leptospira interrogans} (LI), a unique spirochaete bacterium, is self-propelled by rotating the major helix of a super-helicoidal body counter-clockwise and the minor helix clockwise \cite{Goldstein1988a,Charon1984,Nakamura2014}. This super-helix shape only exists near the leading pole, while the other pole takes the form of  a hook. These bacteria are able to swim forwards and backwards and rely on passive diffusion for their rotation. The specific details of how such a shape passively diffuses is thus of interest, let alone how the active swimming would affect  the results.

\begin{figure}
\begin{center}
\includegraphics[width=0.45\textwidth]{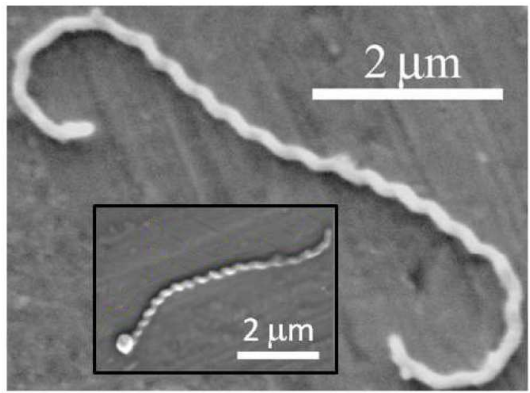}
\caption{A scanning electron microscopy image of \textit{Leptospira interrogans}. The outer image is the standard shape seen. The inset is an example of the straighter shaped bodies that Butenko \textit{et al.} tracked experimentally. Adapted with permission from A. V Butenko, E. Mogilko, L. Amitai, B. Pokroy, and E. Sloutskin, ``Coiled to diffuse: Brownian motion of a helical bacterium'',  \textit{Langmuir}, {\bf 28}, 12941 (2012). Copyright 2012 American Chemical Society.}
\label{SIimage}
\end{center}
\end{figure}

To investigate how a \textit{Leptospira} cell  passively diffuses, Butenko \textit{et al.} employed confocal microscopy to observe its Brownian motion  \cite{Butenko2012}. A fixation process was used to stop all chemical reactions in the body and its motion, reinforce the body shape and prevent its decay. Only bacteria with small or no hook ends were tracked (figure~\ref{SIimage} inset) and the dimensions of the LI were precisely characterised using scanning electron microscopy (figure~\ref{SIimage} and table~\ref{tab:dimensions}). Butenko \textit{et al.} took three-dimensional image stacks every 4.6 seconds from which the diffusion of the helical cell parallel and perpendicular to its major axis were inferred. They also measured the rotational diffusion of the major axis. Comparison with a  previously-existing helical diffusion model \cite{Hoshikawa1979,Hoshikawa1976} showed that it severely underestimated the diffusion coefficients. Attempts to compare the results with the diffusion of a rod were also made \cite{Doi1988a}, with good apparent agreement.  However, the rod model used in \cite{Butenko2012}
 only applies in the exponentially slender limit $1/\log(L/a) \ll 1$, where $L$ is the rod length and $a$ its typical thickness \cite{Lauga2009}. The rod to which the data was  best fit  had $L/a \sim 10^{2}$ meaning $1/\log(L/a) \sim 1/4$. If the actual dimensions of the helical cell are plugged into the exact diffusion coefficient for a prolate spheroid \cite{Chwang2006,Brenner1967}, the agreement with the data turns out to disappear. 

In this paper the Brownian motion of a simple helicoidal model of  LI is considered numerically. A slender-body-theory approach is first employed to determine the hydrodynamics of LI to greater accuracy than the previous approaches \cite{Hoshikawa1979,Hoshikawa1976}. The resulting hydrodynamic properties of the cell are then used to carry out Brownian Dynamics simulations  of the model cell. These results are finally run through a simulated experiment to capture deviations caused by the experimental sampling time before they are compared with the  data of Butenko \textit{et al.} \cite{Butenko2012}. As we detail below, we obtain excellent quantitative agreement with no adjustable parameters. 

The paper is organized as follows. Section~\ref{sec:model} details the four parts of the model: the slender body theory approximation of hydrodynamics in Sec.~\ref{sec:sbt}, the Brownian Dynamics methodology in  Sec.~\ref{sec:bd}, the experiment simulation and sampling method in Sec.~\ref{sec:es} and the geometrical description for the LI helix used in Sec.~\ref{sec:shape}. Section~\ref{sec:results} then presents the results of the paper:  the LI helix hydrodynamics and diffusion in  Sec.~\ref{sec:RFD}, detailed comparison with the  experimental results   in Sec.~\ref{sec:exp}, we then revisit the  prolate-spheroid approximation in Sec.~\ref{sec:ps}, and finish by showing  how our results compare with other models in  Sec.~\ref{sec:om}.

\section{Theoretical model and Numerics} \label{sec:model}

The general motion of a rigid particle subject to thermal fluctuations is described by the Langevin equation 
\begin{eqnarray}
\mathbf{\tilde{m}} \cdot \frac{d \mathbf{\tilde{U}}}{d t} = \sum \mathbf{\tilde{F}} + \mathbf{\tilde{F}}_{Br},
\end{eqnarray}
where $\mathbf{\tilde{m}}$ is the mass/moment of inertia matrix (a 6-by-6 matrix with the mass or moment of inertia of the particle  in the relevant points along the diagonal), $\mathbf{\tilde{U}}$ is a six-component vector containing the linear and angular velocities of the particle, and $\sum  \mathbf{\tilde{F}}$ are the sum of all non-stochastic forces and torques. The above equation assumes that the mass and the moment of inertia of the particle do not change with time. If the body is in a viscous fluid without external forces then $\sum  \mathbf{\tilde{F}}$ is the hydrodynamic force of the fluid on the body, $\mathbf{\tilde{F}}_{H}$.

For $\mu$m-size objects, such as LI bacteria, immersed in a viscous fluid (water or more viscous), the ratio of  inertial forces ($\sim \mathbf{\tilde{m}} \cdot {d \mathbf{\tilde{U}}}/{d t}$) to hydrodynamic forces ($\mathbf{\tilde{F}}_{H}$) is typically small. The equation to solve therefore reduces to that of Brownian Dynamics,  \eref{forcebal}, and the value of $\mathbf{\tilde{F}}_{H}$ must be determined from the instantaneous hydrodynamic response  of the surrounding fluid. 

The ratio of inertial stresses to viscous stresses in the fluid, the Reynolds number, also tends to be small for 
$\mu$m-size objects in a viscous fluid. At low Reynolds numbers, the surrounding fluid is accurately described by the incompressible  Stokes equation \cite{Batchelor1967},
\begin{eqnarray} \label{stokes}
\mu \nabla^{2} \mathbf{u} = \nabla p,\quad \nabla\cdot\mathbf{u} =0,
\end{eqnarray}
 with the no-slip boundary conditions satisfied on the particle surface. In  \eref{stokes}, $\mathbf{u}$ denotes the fluid velocity field, $\mu$  the dynamic viscosity, and $p$ is the pressure. As the equations in  \eref{stokes} are linear and time independent, the instantaneous forces and torques on any submerged body are linearly related to the linear and angular velocity of the body as
\begin{eqnarray} \label{resistance}
\mathbf{\tilde{F}}_{H}  = \mathbf{R}_{\tilde{F} \tilde{U}} \mathbf{\tilde{U}},
\end{eqnarray} 
where $\mathbf{R}_{\tilde{F} \tilde{U}}$ is  the symmetric  resistance matrix, proportional to the  viscosity of the fluid and function of the size and  shape of the particle. Substituting  \eref{resistance} into  \eref{forcebal} gives
\begin{eqnarray} \label{langevin}
\mathbf{\tilde{U}} = \mathbf{R}_{\tilde{F} \tilde{U}}^{-1}  \mathbf{\tilde{F}}_{Br},
\end{eqnarray} 
where the change of sign has been absorbed into $\mathbf{\tilde{F}}_{Br}$. In this formulation $\mathbf{\tilde{F}}_{Br}$ is assumed to be white noise with the statistical properties of  \eref{Fmean} and  \eref{Fcovars}.

Equations~\eref{Fmean},  \eref{Fcovars}, and 
\eref{langevin}  describe the velocity of a rigid particle in a specific frame moving with the body called the centre of mobility which is unique \cite{Brenner1967}.  In this frame the sub-matrix of $\mathbf{R}_{\tilde{F} \tilde{U}}$ that relates force and rotation, $\mathbf{R}_{F \Omega}$, is symmetric. To solve for the rotational and translational motion of that frame in the  laboratory frame one needs to integrate in time
\begin{eqnarray}
\frac{d \mathbf{e_{i}}}{d t} &=& \mathbf{\Omega}\times\mathbf{e_{i}}, \label{rotation}\\ 
\frac{d \mathbf{r}}{d t} &=& \mathbf{U}, \label{translation}
\end{eqnarray} 
where $\mathbf{e_{i}}$ represents the direction of one of the body-frame basis vectors ($\mathbf{e_{1}}$, $\mathbf{e_{2}}$ or $\mathbf{e_{3}}$), $\mathbf{\Omega}$ is the angular velocity of the body, $\mathbf{r}$ its position and $\mathbf{U}$ is the velocity vector. The location of the centre of mobility frame, $\mathbf{R'}$, relative to another fixed frame on the body, $\mathbf{R}$, is given by
\begin{eqnarray}\label{cm}
[\mathbf{R}_{M \Omega} - \mathbf{I} Tr(\mathbf{R}_{M \Omega})] \cdot (\mathbf{R'} -\mathbf{R}) = \epsilon_{ijk} R_{F \Omega; j k} \mathbf{e_{i}},
\end{eqnarray}
where $\mathbf{I}$ is a 3-by-3 identity matrix, $Tr(\cdot)$ indicates the trace, $\epsilon_{ijk}$ is the Levi-Civita permutation tensor, and $\mathbf{R}_{\tilde{F} \tilde{U}}$ has been divided into four  3-by-3 matrices such that 
\begin{eqnarray}
\mathbf{R}_{\tilde{F} \tilde{U}} = \left(\begin{array}{ c c}
\mathbf{R}_{F U} & \mathbf{R}_{F \Omega} \\
\mathbf{R}_{F \Omega}^{T} & \mathbf{R}_{M \Omega}
\end{array} \right).
\end{eqnarray}
In the above equation repeated indices are summed over from 1 to 3 and all values are described in the $\mathbf{R}$ frame. 
 
In summary, \eref{Fmean}, \eref{Fcovars}, \eref{langevin}, \eref{rotation}, and \eref{translation}  fully describe the behaviour of one rigid particle subject to thermal fluctuations and the desired statistics may then be obtained from  multiple realizations of these equations. The resistance matrix, $\mathbf{R}_{\tilde{F} \tilde{U}}$, of the particle must therefore be known in order to carry out these calculations. Generally $\mathbf{R}_{\tilde{F} \tilde{U}}$ is not known exactly, and approximations must be made. Below, we show how to obtain an approximate value for  $\mathbf{R}_{\tilde{F} \tilde{U}}$ using so-called slender body theory. We then solve numerically  \eref{Fmean}, \eref{Fcovars}, \eref{langevin}, \eref{rotation}, and \eref{translation}.
Finally, due of the long times between each experimental measurement in  Butenko \textit{et al.}'s investigation, the experimentally-determined diffusion coefficients  deviate from their exact values. We thus carry out a simulation of their experimental procedure  in order to enable an accurate comparison of our numerical results with the experiment.

\subsection{Slender body theory of hydrodynamics} \label{sec:sbt}

For rigid bodies which are much longer than they are thick, the resistance tensor $\mathbf{R}_{\tilde{F} \tilde{U}}$ can be approximated to good accuracy by an asymptotic method called slender body theory. This method approximates the flow around a slender object at low Reynolds number by placing a series of force and source dipole singularities along the body centerline \cite{1976,Johnson1979}. The strengths of these singularities are expanded in powers of the ratio between the body thickness and its length and matched to the boundary conditions at the  surface of the body. The resulting equations form a set of integral equations for the force distribution along the body in response to a given motion \cite{1976,Johnson1979,Gotz2000}. Historically, there has been two major formulations of slender body theory, Lighthill's  \cite{1976} and Johnson's  \cite{Johnson1979}. In Lighthill's slender body theory, the cross sectional shape of the body is held constant and end effects are ignored. Alternatively Johnson's more accurate slender body theory included end effects and the possibility for changing cross sectional thickness provided the cross sectional thickness behaved like a prolate spheroid near the ends. In this paper we use of Johnson's slender body theory, which describes the hydrodynamics with relative accuracy of $\epsilon^2\ln \epsilon$ when $\epsilon$ is the typical aspect ratio of the slender body. 

The integral equation in Johnson's slender body theory is given by
\begin{eqnarray} \label{SBT}
8 \pi \mu \frac{\partial \mathbf{x}(s,t)}{\partial t} = \lambda[\mathbf{f}] +K_{a}[\mathbf{f}] +K_{b}[\mathbf{f}],
\end{eqnarray}
where $\mathbf{x}(s,t)$ is the location  of the body's centerline parametrized by arc length, $s$, and $\mathbf{f}(s)$ is the force distribution along the centerline of the body. The operators in  \eref{SBT} are given by
\begin{eqnarray}
\lambda[\mathbf{f}] &=& \left[d\left(\mathbf{I}+ \mathbf{\hat{t} \otimes \hat{t}}\right)+2\left(\mathbf{I}- \mathbf{\hat{t} \otimes \hat{t}}\right)\right] \cdot \mathbf{f}(s), \\
\label{Ka}K_{a}[\mathbf{f}] &=& \left(\mathbf{I}+ \mathbf{\hat{t} \otimes \hat{t}}\right) \cdot \int_{-l}^{l} \frac{\mathbf{f}(s') -\mathbf{f}(s)}{|s'-s|} \,d s', \\
K_{b}[\mathbf{f}] &=& \int_{-l}^{l} \left[ \frac{\mathbf{I}+ \mathbf{\hat{R}} \otimes \mathbf{\hat{R}}}{|\mathbf{R}|}-\frac{\mathbf{I}+ \mathbf{\hat{t} \otimes \hat{t}}}{|s'-s|}\right] \cdot \mathbf{f}(s') \,d s', \label{kb}
\end{eqnarray}
where $d = -\log(\epsilon^{2} e)$, $\mathbf{\hat{t}}$ is the tangent vector to the centerline at $s$, $\mathbf{R} = \mathbf{x}(s)-\mathbf{x}(s')$, $\mathbf{\hat{R}} = \mathbf{R}/|\mathbf{R}|$, $\epsilon = r_{b}/2l$ is the slenderness parameter, $2 r_{b}$ is the thickness of the body cross section, $2l$ is the centerline length of the body and $e$ is $\exp(1)$. These equations assume the cross sectional radius of the object is $r_{b}\sqrt{1-s^{2}}$. In what follows we use a dimensionless framework with $l =\mu =1$ unless otherwise stated. 

Equation~\eref{SBT} provides the link between the motion of a slender body  and the force distributed along its centerline. If this force distribution can be determined,  the total force required to create a given motion, and by extension the resistance coefficient, can then be determined. Solving for the force distribution is done numerically using a  Galerkin method \cite{KendallE.Atkinson2009} and  writing the force density as a finite sum 
\begin{eqnarray} \label{forceexpand}
\mathbf{f}(s) = \sum_{n=0}^{N} \mathbf{a}_{n} P_{n}(s),
\end{eqnarray}
where $P_{n}(s)$ are Legendre polynomials of order $n$, $\mathbf{a}_{n}$ are vectors representing the Legendre polynomial coefficients in the three Cartesian directions and $N$ is the summation  order. In the limit of $N\rightarrow\infty$ the numerical approximation becomes exact due to the orthogonality of the Legendre polynomials. The specific expansion in Legendre polynomials is used here  because $P_{n}(s)$ are eigenfunctions of the integral in $K_{a}[]$ (\ref{Ka}), with eigenvalues $-L_{n}$ \cite{Gotz2000},
\begin{eqnarray}
K_{a}[P_{n} \mathbf{A}] = - \left(\mathbf{I}+ \mathbf{\hat{t} \hat{t}}\right) \cdot \mathbf{A} L_{n} P_{n},
\end{eqnarray}
where $L_{0} = 0$, $L_{n} = 2 \sum_{i=0}^{n} 1/i$ for $n>0$, and $\mathbf{A}$ is an arbitrary matrix.

Using the decomposition  \eref{forceexpand} allows to reduce 
  \eref{SBT}  into a series of linear equations for the $\mathbf{a}_{n}$ vectors. These equations are found by multiplying  \eref{SBT} by $P_{m}(s)$ and integrating $s$ over [-1,1], leading to
\begin{eqnarray} \label{SBT2}
\fl 8 \pi \int_{-1}^{1} P_{m}(s)\frac{\partial \mathbf{x}(s,t)}{\partial t} \,ds    = \sum_{n=0}^{\infty} \left\{ \int_{-1}^{1} P_{m}(s) (\lambda[\mathbf{I} P_{n}]+K_{a}[\mathbf{I}P_{n}]+K_{b}[\mathbf{I}P_{n}])  \,ds\right\} \cdot\mathbf{a}_{n} .
\end{eqnarray}
The integrals on the left hand side (LHS) and right hand side (RHS) of \eref{SBT2} only depend on the shape and motion of the body and its centerline. Therefore these integrals can be determined for a given set of Legendre polynomials. 
The LHS integral and the integrals of $P_{m}(s)\lambda[\mathbf{I} P_{n}]$ and $P_{m}(s)K_{a}[\mathbf{I}P_{n}]$ can be simply evaluated using MATLAB  \cite{MATLAB2014}. The integral of $P_{m}(s)K_{b}[\mathbf{I} P_{n}]$ is not as simple to evaluate. Analytically $K_{b}[]$, \eref{kb}, is non-singular at $s=s'$ but the individual terms in $K_{b}[]$ are. This can be checked by performing the Taylor series of $K_{b}[]$ around $s=s'$. The singular nature of the individual terms causes the direct numerical sampling of $s=s'$ to give non-numerical values. Therefore simple numerical integration schemes will not work. The quadrature integration methods in MATLAB \cite{MATLAB2014} were used to overcome this. Quadrature methods handle the sampling issue by using the points around $s=s'$ to determine the limiting behaviour of the integral at $s=s'$. The MATLAB quadrature methods work best when the singular points are at the boundaries. Therefore the integral is divided into two regions $s \in [-1,1], s' \in [-1,s]$ and $s \in [-1,1], s' \in [s,1]$ to assist with the calculation.

We note that slender body theory does not include the local torque due to surface rotation along the centerline of the elongated body. This was added  to the  system by adding  a series of rotlet singularities \cite{Chwang2006} to the body's centerline. Expanding the rotlet singularities in powers of the slenderness parameter, the local contribution of the rotlets to the flow is $\gamma(s) =16 \pi \epsilon^{2} (1-s^{2})( \mathbf{\Omega}\cdot \mathbf{\hat{t}} )\mathbf{\hat{t}}$, where $\gamma(s)$ is the rotlet strength at $s$. The total additional torque from surface rotation was then obtained by integrating $\gamma$ over $s$ using MATLAB. 

With the method outlined above, the  coefficients for  $\mathbf{a}_{n}$ and $\gamma(s)$ were obtained numerically, and thus the matrix 
$\mathbf{R}_{\tilde{F} \tilde{U}}$ was constructed  by calculating the total force and torque created from translation or rotation in a single direction. The value of the truncation order $N$ was chosen such that the change in the values of $\mathbf{R}_{\tilde{F} \tilde{U}}$ between the $N-2$ and $N$ cases is less than three decimal places.

\begin{figure}
\begin{center}
\includegraphics[width=0.9\textwidth]{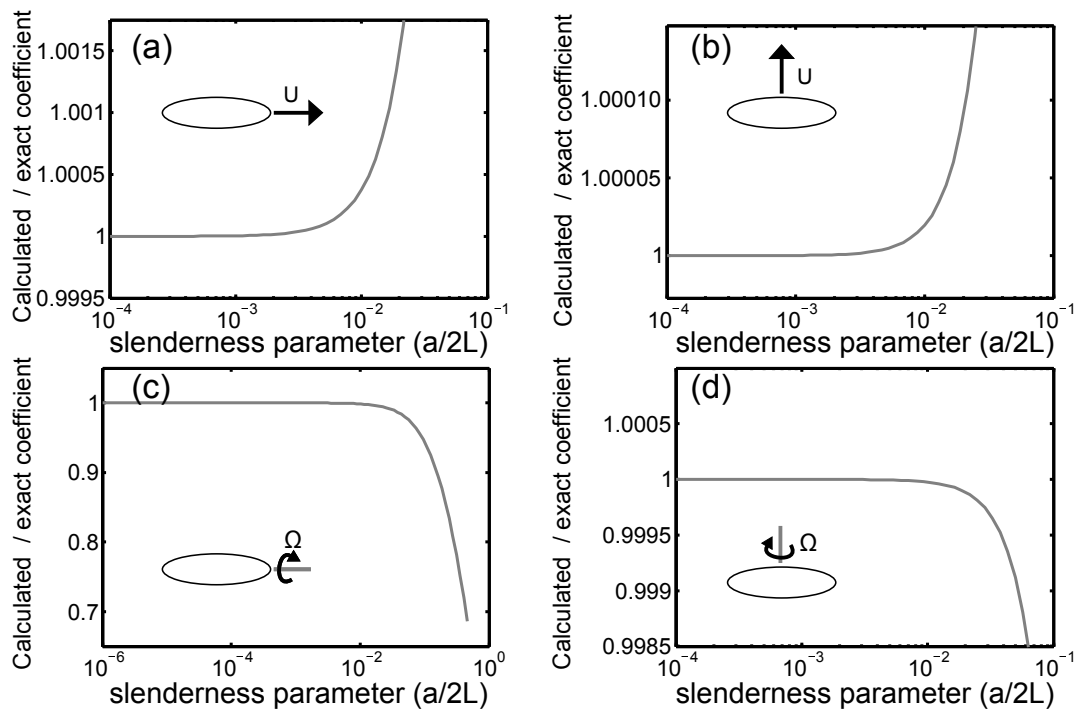}
\caption{Ratio of resistance coefficients calculated using slender body theory to their exact values \cite{Chwang2006} for a prolate spheroid;  (a):  coefficient that relates the drag force parallel to the prolate spheroid's major axis from motion in the same direction, $C_{\Vert}$; (b):  coefficient that relates the drag perpendicular to the major axis from motion in the same direction, $C_{\perp}$; (c):  coefficient that relates the torque parallel from rotation in the same direction, $k_{\Vert}$; (d): coefficient that relates the torque perpendicular from rotation in the same direction, $k_{\perp}$.  Results are shown for $N=4$ while higher values of  $N$ show no discernible changes.}
\label{roderror}
\end{center}
\end{figure}

The accuracy of our implementation of  slender body theory was tested  by comparing the results from \eref{SBT2} to the exact  resistance coefficients for a prolate spheroid \cite{Chwang2006}. Figure~\ref{roderror} shows the ratio between our numerical results  and the  exact solutions for different values of the slenderness parameter, $\epsilon$. All ratios are seen to converge to 1 as the slenderness decreases ($\epsilon$ goes to 0). The slowest term to converge is the coefficient that relates torque parallel to the major axis to the rotation in the same direction, $k_{\Vert}$, which is calculated from the added rotlets. As rotlets create flows which decay as r$^{-2}$,  the error could indicate that the rotlets are reaching a strength where interactions not taken into account in the value of $\gamma(s)$ begin to influence for larger values of $\epsilon$. The results of the slender body theory program were also tested against all the cases in Johnson's original article on his slender body theory \cite{Johnson1979} with excellent agreement (not shown).

\subsection{Brownian Dynamics} \label{sec:bd}

With all coefficients in $\mathbf{R}_{\tilde{F} \tilde{U}}$  found,  \eref{Fmean}, \eref{Fcovars}, \eref{langevin}, \eref{rotation}, and \eref{translation} can be solved. This is done by first transforming $\mathbf{R}_{\tilde{F} \tilde{U}}$ into the centre of mobility frame \cite{Brenner1967} and then solving the equations with an Euler-Maruyama method. The Euler-Maruyama method approximates the trajectories that the particles take (a strong integrator) and converges to the exact solution with the properties $\left\langle(X(t) - Y(t))^{2} \right\rangle \leq C \Delta t$ (an order 0.5 accurate method) where $X(t)$ is the exact solution, $Y(t)$ is the approximation, $C$ is a constant and $\Delta t$ is the time between two consecutive time steps \cite{Kloeden1992}. In the Euler-Maruyama method  \eref{Fmean}, \eref{Fcovars}, \eref{langevin}, \eref{rotation}, and \eref{translation} become
\begin{eqnarray}
\mathbf{\tilde{U}} \Delta t &=& \mathbf{R}_{\tilde{F} \tilde{U}}^{-1}  \mathbf{\tilde{F}}_{Br}\Delta t,\\
\left\langle\mathbf{\tilde{F}}_{Br}\Delta t\right\rangle &=& 0,  \\
\left\langle\mathbf{\tilde{F}}_{Br} \Delta t \otimes  \mathbf{\tilde{F}}_{Br}\Delta t\right\rangle &=& 2 k_{b} T \mathbf{R}_{\tilde{F} \tilde{U}} \Delta t, \\
\mathbf{e_{i}}(t_{n+1})& =& \mathbf{e_{i}}(t_{n}) + \mathbf{\Omega}(t_{n}) \Delta t \times \mathbf{e_{i}}(t_{n}) , \label{roteuler} \\
 \mathbf{r}(t_{n+1}) &=& \mathbf{r}(t_{n}) + \mathbf{U}(t_{n}) \Delta t, 
\end{eqnarray}
where $t_{n}$ denotes the $n$th time step. Computationally,  $\mathbf{\tilde{F}}_{Br} \Delta t$ is taken to be Gaussian and is generated by the MATLAB function \textit{mvnrnd}. Care is needed with \eref{roteuler}, as it may not conserve vector length or the orthogonality of the three body basis vectors $\mathbf{e_1}$, $\mathbf{e_2}$ and $\mathbf{e_3}$. This is overcome by determining only $\mathbf{e_1}(t_{n+1})$ and $\mathbf{e_2}(t_{n+1})$ using  \eref{roteuler}; $\mathbf{e_1}(t_{n+1})$ is renormalized and any projection of $\mathbf{e_2}(t_{n+1})$ on the renormalized $\mathbf{e_1}(t_{n+1})$ is removed from $\mathbf{e_2}(t_{n+1})$ before it too is renormalized. Finally the third vector $\mathbf{e_3}(t_{n+1})$ is determined by taking the cross product of $\mathbf{e_1}(t_{n+1})$ and $\mathbf{e_2}(t_{n+1})$. The statistical behaviour of the diffusing particle is determined by solving the above equations $q$ times from $t=0$ to an end time, $t_{fin}$, with a set value of $\Delta t$. 

\begin{figure}
\begin{center}
\includegraphics[width=0.9\textwidth]{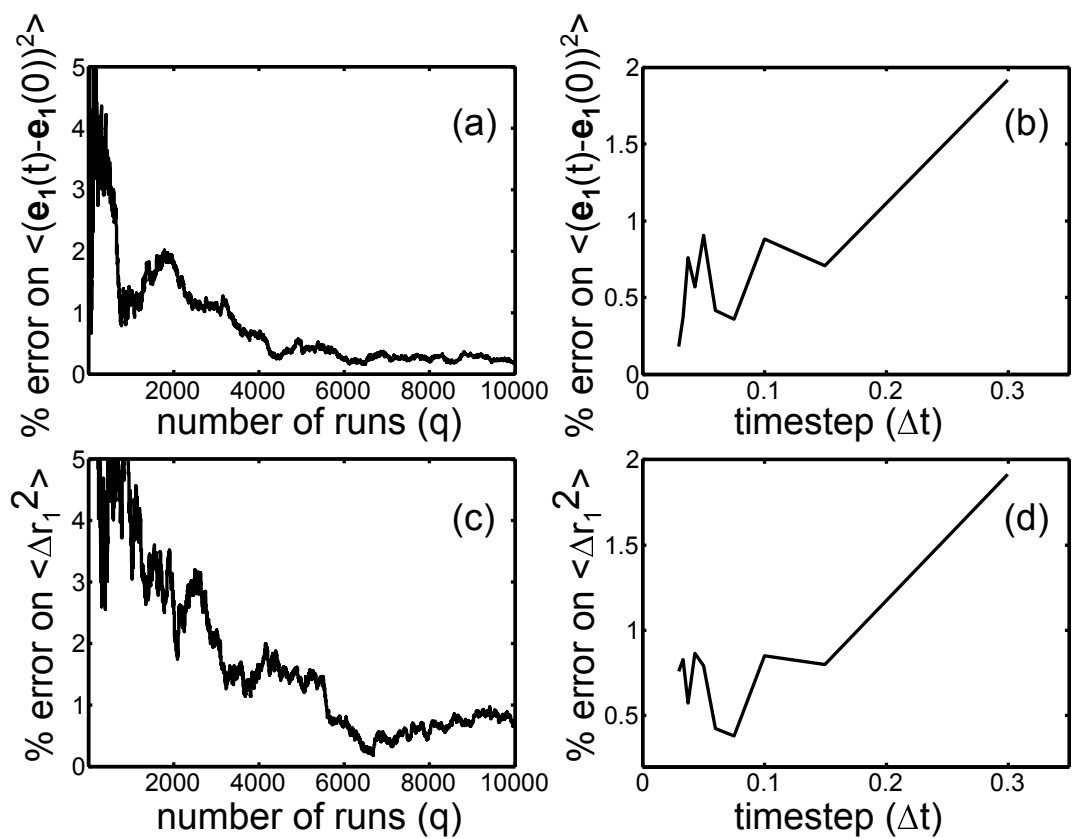} 
\caption{Brownian Dynamics for the diffusion of a sphere. Percentage error from the numerical calculation of $\left\langle(\mathbf{e_{1}}(t)-\mathbf{e_{1}}(0))^{2}\right\rangle$, (a-b), and the mean squared displacement, $\left\langle\Delta r_{1}^{2}\right\rangle$, (c-d),  at $t_{fin} =30$ (scaled units); (a) and (c) show the error due to changing number of runs ($q$) when $\Delta t =0.03$, while (b) and (d) show the error for different time steps $\Delta t$ with $q=10000$.}
\label{spheretest}
\end{center}
\end{figure}

Our numerical implementation of  the Brownian Dynamics  method  was tested by investigating the statistical behaviour of a diffusing sphere and of a prolate spheroid. In both cases the mean squared displacement $\left\langle \Delta r_{i}^{2}\right\rangle$ along separate directions in the lab frame and the rotational diffusion of $\mathbf{e_{i}}$ in the form of $\left\langle(\mathbf{e_{i}}(t)-\mathbf{e_{i}}(0))^{2}\right\rangle$ was considered. 
A diffusing sphere follows   $\left\langle\Delta r_{i}^{2}\right\rangle = 2Dt$ along the $i$th direction and $\left\langle(\mathbf{e_{i}}(t)-\mathbf{e_{i}}(0))^{2}\right\rangle = 2[1-\exp(-2D_{\theta}t)]$, where $D = k_{b} T / 6 \pi \mu a$ and $D_{\theta} = k_{b} T / 8 \pi \mu a^{3}$. The percentage difference between the analytic formula above and the numerics was determined at time $t_{fin}=30$ scaled units for various $q$ and $\Delta t$. Figure~\ref{spheretest} shows the results of these tests for both $\left\langle\Delta r_{1}^{2}\right\rangle$ and $\left\langle(\mathbf{e_{1}}(t)-\mathbf{e_{1}}(0))^{2}\right\rangle$. The erratic behaviour seen is due to the stochastic nature of the integration program. The error is seen to decrease with increasing $q$ and decreasing $\Delta t$. For $q=10000$ and $\Delta t =0.03$ (scaled units) the error at $t_{fin}=30$ (scaled units) is less than 1\%, and these are the values used for the diffusion calculation of LI in Sec.~\ref{sec:results}. 

\begin{figure}
\begin{center}
\includegraphics[width=0.9\textwidth]{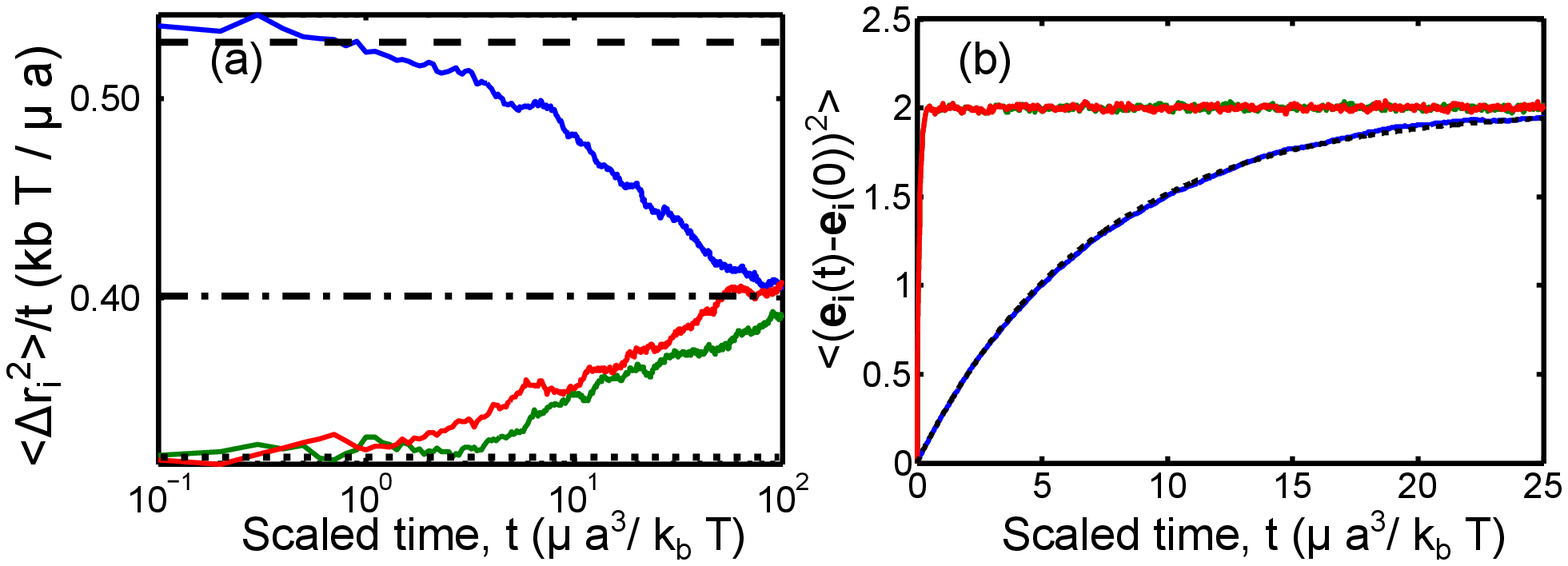}
\caption{(Colour online) Brownian Dynamics for the diffusion of a prolate spheroid.  Directional mean squared displacements divided by time and non-dimensionalized (a) and $\left\langle(\mathbf{e_{i}}(t) -\mathbf{e_{i}}(0))^{2}\right\rangle$ (b). In (a): the solid blue line is mean squared displacement along the (1,0,0) direction, the solid green is for (0,1,0), the solid red is (0,0,1). The black straight lines correspond to diffusion purely along the major axis (dashed), minor axis (dotted) and the long-time diffusion constant obtained when orientations are lost (dash-dotted). 
In (b) the black dashed line is the theoretical prediction of how $\mathbf{e_1}$ (so $i=1$) should change, the solid blue line is the calculated change in $\mathbf{e_1}$, the solid green is $\mathbf{e_2}$ and the solid red is $\mathbf{e_3}$. The results were obtained with $q=10000$, $t_{fin}$ = 25, and $\Delta t$ = 0.03 scaled units. The prolate spheroid had an aspect ratio of 55.6. All lengths are scaled by half the length of the major axis, $a$, and  times by  $\mu a^{3}/k_{b} T$.}
\label{prolatetest}
\end{center}
\end{figure}

For a prolate spheroid with major axis $\mathbf{e_1}$ initially aligned with the direction (1,0,0), diffusion along and against (1,0,0) starts like $\left\langle\Delta r_{\Vert}^{2}\right\rangle = 2D_{\Vert}t$ and $\left\langle\Delta r_{\perp}^{2}\right\rangle = 2D_{\perp}t$, respectively, where $D_{\Vert}=k_{b} T / C_{\Vert}$, $D_{\perp}=k_{b} T / C_{\perp}$, and $C_{\Vert}$ (resp.~$C_{\perp}$ )  is the coefficient that relates the drag force parallel (resp.~perpendicular) to the prolate spheroid's major axis from motion in the same direction. As time progresses orientation is lost and the mean squared displacement along these directions become $\left\langle\Delta r_{i}^{2}\right\rangle = 2(D_{\Vert} +2D_{\perp}) t/3$. Similarly to above, we expect $\left\langle(\mathbf{e_1}(t)-\mathbf{e_1}(0))^{2}\right\rangle = 2[1-\exp(-2D_{\theta,1}t)]$, where $D_{\theta,1} = k_{b} T / k_{\perp}$ and $k_{\perp}$ is the coefficient that relates the torque perpendicular to the major axis from rotation in the same direction. The rotational diffusion of $\mathbf{e_2}$ and $\mathbf{e_3}$ is very rapid for a prolate spheroid, and is not relevant on the time scales considered. Therefore we only consider the rotational diffusion caused by rotations around the minor axes. In figure~\ref{prolatetest} we display the directional mean square displacement and $\left\langle(\mathbf{e_{i}}(t)-\mathbf{e_{i}}(0))^{2}\right\rangle$ for a prolate spheroid. The black dashed lines are  theoretical predictions while the solid lines are the numerical predictions, and we obtain excellent quantitative  agreement.

\subsection{Numerical simulation of the experiment} \label{sec:es}

 Modern three-dimensional real-space microscopy techniques have trouble determining the three dimensional diffusion of micron-sized objects in water with sufficient speed to get the full detail of Brownian Dynamics simulations. Though fast CCD and CMOS cameras are available, confocal laser scanning (at sufficient digital resolution, such as 512x512 pixels) is typically limited to 30 frames per second; moreover, many images at different focal planes are needed to obtain the orientation of the body and its location in a three-body environment.  Hence a three-dimensional image can take seconds to obtain. The determination of the diffusion characteristics therefore must be inferred from the limited measurements available. Butenko et al.'s experimental setup made a three-dimensional measurement of 50 focal planes every 4.6 seconds with a 30 fps resonant laser scanning system. The diffusive behaviour was estimated from the results. In each measurement, the authors determined the direction of the major axis, $\mathbf{e_1}$, and the position of the cell; the value of  $\left\langle(\mathbf{e_1}(t) -\mathbf{e_1}(0))^{2}\right\rangle$ was then obtained directly from the  measurements for $\mathbf{e_1}$. The displacement parallel and perpendicular to the major axis was approximated from the change in the position between two measurements, at times $t_{n}$ and $t_{n-1}$, by dividing the change into motion parallel and perpendicular to $\mathbf{e_1}(t_{n})$. The steps parallel (perpendicular) to $\mathbf{e_1}$ were than summed over a time period, $t_{period}$, starting at different time steps, $\tau$, where $\tau \in (0,t_{fin}-t_{period})$ and $t_{fin}$ is the length of the experiment. This gives a list of the displacement parallel (perpendicular) in the time periods $\tau \rightarrow \tau +t_{period}$, for different starting time steps $\tau$. The mean squared displacement parallel (perpendicular) after a time $t_{period}$ is then given by averaging over the squared values of the list.
 
 The presence of long times between measurements prompts us to apply the same procedure to the Brownian Dynamics  data in order  to capture any sampling effects this may have on the results. We thus determined the orientation and centre of mobility of the body every 4.6 seconds from our numerical simulations. The value of $\left\langle(\mathbf{e_1}(t) -\mathbf{e_1}(0))^{2}\right\rangle$, the mean squared displacement parallel to the major axis, $\left\langle\Delta r_{\Vert}^{2}\right\rangle$, and the mean squared displacement perpendicular to the major axis, $\left\langle\Delta r_{\perp}^{2}\right\rangle$, were then calculated as described above. To estimate the diffusion coefficients, the  behaviour of $\left\langle\Delta r_{\Vert}^{2} \right\rangle$ and $\left\langle\Delta r_{\perp}^{2}\right\rangle$ was assumed to be of the form $2Dt+c$ while that of  $\left\langle(\mathbf{e_1}(t) -\mathbf{e_1}(0))^{2}\right\rangle$ was assumed to be of the form $2[1-\exp(-2Dt)]$. We then used least-square regressions  to find the optimal `diffusion coefficients', $D$, and intercepts, $c$, for the processed numerical Brownian Dynamics data.

\begin{figure}
\begin{center}
\includegraphics[width=0.8\textwidth]{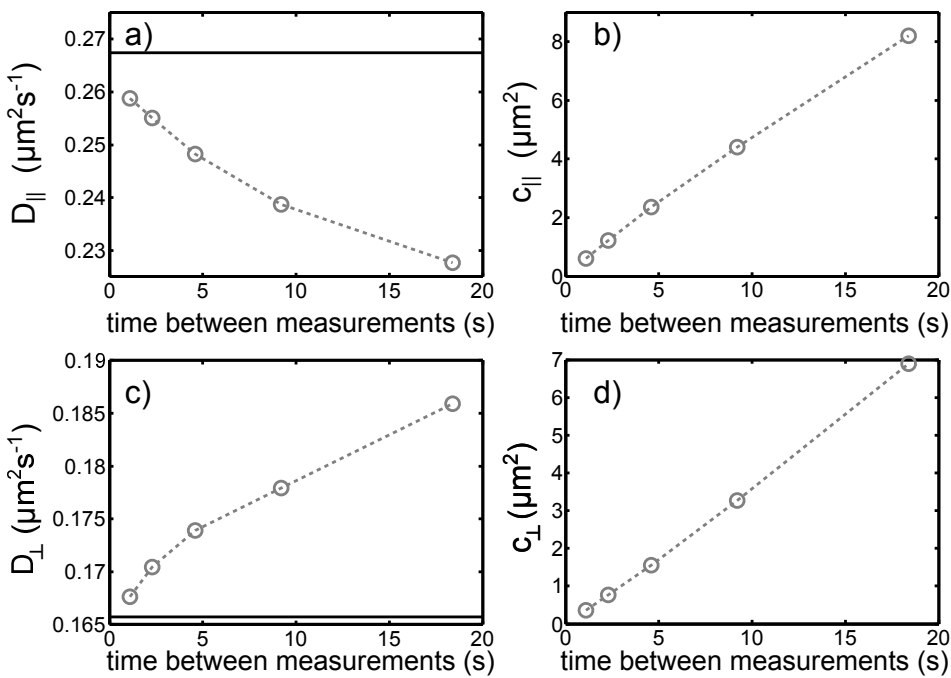} 
\caption{Diffusion coefficients ($D$) and intercepts ($c$) predicted by the experimental sampling simulation for the diffusion parallel [(a): $D_{\Vert}$;  (b): $c_{\Vert}$] and perpendicular [(c): $D_{\perp}$; $c_{\perp}$] to the major axis of a prolate spheroid with semi-major and semi-minor axis lengths of  $4.55~\mu\mbox{m}$ and $0.0695~\mu\mbox{m}$, respectively (dimensions  similar to those of LI; $k_{b} T /\mu = 3.70~\mu\mbox{m}^{3} s^{-1}$). The circles and the dashed lines are the results of  experimental processing (least squared regression) of the Brownian Dynamics data while the solid lines are the theoretical diffusion coefficients with $c=0$. 
}
\label{experimentsim}
\end{center}
\end{figure}

Figure~\ref{experimentsim} shows the effect of the experimental  methodology on the predicted diffusion coefficients and intercepts obtained from the mean squared displacement data in the case of a prolate spheroid. The results show, as could have been intuitively guessed,  that the processing has a quantitative effect on the predicted behaviour and should thus be included to the Brownian Dynamics simulation in order to best replicate the behaviour observed. As the time between measurements gets larger, the predicted  value for $D_{\Vert}$ further underestimates the exact solution while that of  $D_{\perp}$  overestimates it. Similarly, the intercept of the least squared regression tends to increase with the time between the measurements. This is caused by the rotation of the body. Indeed, in the time between observations the body has rotated by a finite amount. Therefore not all of the motion assumed to be parallel to the major axis was actually parallel to it. Since motion perpendicular to the major axis has a greater resistance than in the  parallel direction, the observed net displacement parallel (resp.~perpendicular) decreases (resp.~increases) compared to the actual displacement and the diffusion coefficient parallel (resp.~perpendicular) to the major axis is underestimated (resp.~overestimated).

\subsection{Geometrical description of \textit{Leptospira interrogans}} \label{sec:shape}

\begin{table}
\caption{The dimensions of LI found by Butenko \textit{et al.} \cite{Butenko2012}: $L$ is the length of the helix measured along the major axis, $r_{h}$  the helix radius, $P$  the helix pitch, $2 r_{b}$ is the thickness of the helix cross section. Each dimension is assumed to be normally distributed.}
\begin{indented}
\item[] \begin{tabular}{@{} l l l} \br
Value & Mean~($\mu$m)  & Stand.~dev.~($\mu$m) \\ \mr
$L$ & 9.10 & 4 \\
$r_{h}$ & 0.0850 & 0.03 \\
$P$ & 0.392 & 0.07 \\
$r_{b}$ & 0.0695 & 0.02 \\ \br
\end{tabular}
\label{tab:dimensions}
\end{indented}
\end{table}

\begin{figure}[b]
\begin{center}
\includegraphics[width=0.8\textwidth]{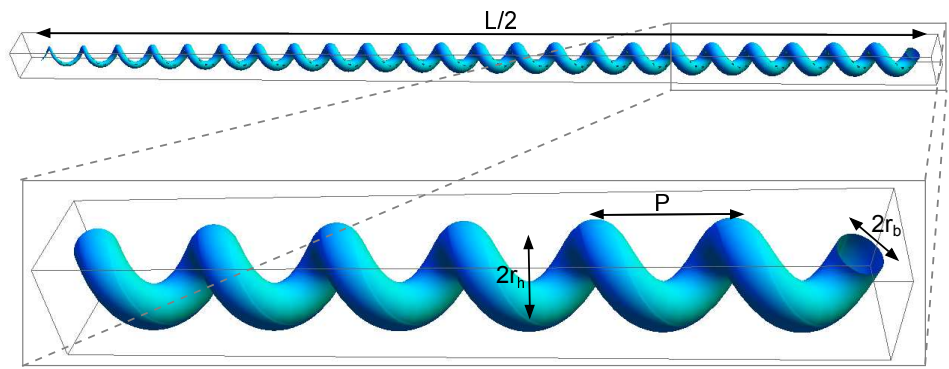} 
\caption{(Colour online) A visual representation of half the LI helix used by the slender body program. $L$ is the length measure along the helix axis, $P$ is the helix pitch, $r_{h}$ is the helix radius and $2 r_{b}$ is the thickness of the body cross section at the centre. }
\label{helixshape}
\end{center}
\end{figure}

With the  method outlined above,  a numerical simulation of the experiment can be performed for any slender shape. In order  to accurately  simulate Butenko \textit{et al.}'s experiment, a suitable geometric form is needed to describe \textit{Leptospira interrogans}. The typical form of a cell tracked in the experiments is shown in figure~\ref{SIimage}. Assuming all the bodies tracked were like the case in figure~\ref{SIimage}, it is safe to ignore the hook ends of LI and treat the cell as a straight helix with a centerline described as
\begin{eqnarray} \label{centreline}
\mathbf{x(s)} = \left( \frac{\theta}{k}, r_{h} \cos \theta, r_{h} \sin \theta \right)
\end{eqnarray}
where $r_{h}$ is the helix radius, $k=2 \pi /P$, $\theta /k = s/\alpha$, $\alpha^{2} = 1 + r_{h}^{2} k^{2}$, $P$ is the helix pitch. The dimensions of this helix were taken to be the mean dimensions of LI listed in table~\ref{tab:dimensions}. We further assume that the cross sectional radius of the body  behaves like an ellipse, and is thus given by $r_{b} \sqrt{1-s^{2}}$. Johnson showed \cite{Johnson1979} that the choice of cross section has little effect on the total force and torque felt by a helical body and so justifies this assumption. We show in figure~\ref{helixshape} half the body of the LI helix and illustrates each of the parameters.

\section{Results and discussion} \label{sec:results}

\subsection{Resistance matrix and diffusion of \textit{Leptospira interrogans}} \label{sec:RFD}

 Using the geometric model of LI from Sec.~\ref{sec:shape} in the slender body program summarized in Sec.~\ref{sec:sbt}, we can determine the resistance matrix of LI in the center of the  centerline of the cell  as
\begin{eqnarray} \label{RM}
\fl \mathbf{R}_{\tilde{F} \tilde{U}}/\mu = \left(
\begin{array}{c c c c c c}
   15.170&   -0.007&    0.008&    0.041&   -0.169&    0.211\\
   -0.007&   23.274&    0.005&    0.009&   -0.019&   -0.036\\
    0.008&    0.005&   23.271&   -0.012&   -0.036&   -0.002\\
    0.041&    0.009&   -0.012&    0.275&   -0.058&    0.072\\
   -0.169&   -0.019&   -0.036&   -0.058&  205.969&   -0.234\\
    0.211&   -0.036&   -0.002&    0.072&   -0.234&  206.073
\end{array}\right).
\end{eqnarray}
The above matrix is obtained with $N=34$, is accurate to three decimal places,  and all lengths are in $\mu$m. We note that it is dominated by the diagonal terms and only has weak couplings off the diagonal. As the top-right and bottom-left sub matrices in \eref{RM} are not symmetric, the matrix is obviously not expressed in the centre of mobility frame. Using \eref{cm} we obtain that the centre of mobility is offset from (0,0,0) in \eref{centreline} by (0,0.001,0.001). Rewriting \eref{RM} in this frame produces no discernible changes to the numerical results presented below.
 
With the value of $\mathbf{R}_{\tilde{F} \tilde{U}}$ known,  we can use it in our  Brownian Dynamics computation (Sec.~\ref{sec:bd}) to determine the diffusion coefficients of LI. From this computation, we predict the true diffusion coefficients parallel, perpendicular and the rotational diffusion coefficient against the major axis of the LI helix to be
\begin{eqnarray}
D_{\Vert} &=& 0.244~\mu \mbox{m}^{2} s^{-1}, \label{dpara} \\ 
D_{\perp} &=& 0.160~\mu \mbox{m}^{2} s^{-1}, \label{dperp}\\ 
D_{\theta} &=& 0.0180~\mbox{rad}^{2} s^{-1}, \label{drot}
\end{eqnarray}
where $k_{b} T /\mu = 3.70~\mu\mbox{m}^{3} s^{-1}$ (as in Butenko \textit{et al.}'s experiment \cite{Butenko2012}), $q=10000$, $t_{fin}=100s$, and $\Delta t = 0.1 s$. The values show relatively fast diffusion along the bodies major axis and slow rotational diffusion against the helix's major axis (diffusion from rotations around the minor axes) as would be expected.

\subsection{Comparison between experiment and numerics} \label{sec:exp}

\begin{figure}
\begin{center}
\includegraphics[width=0.9\textwidth]{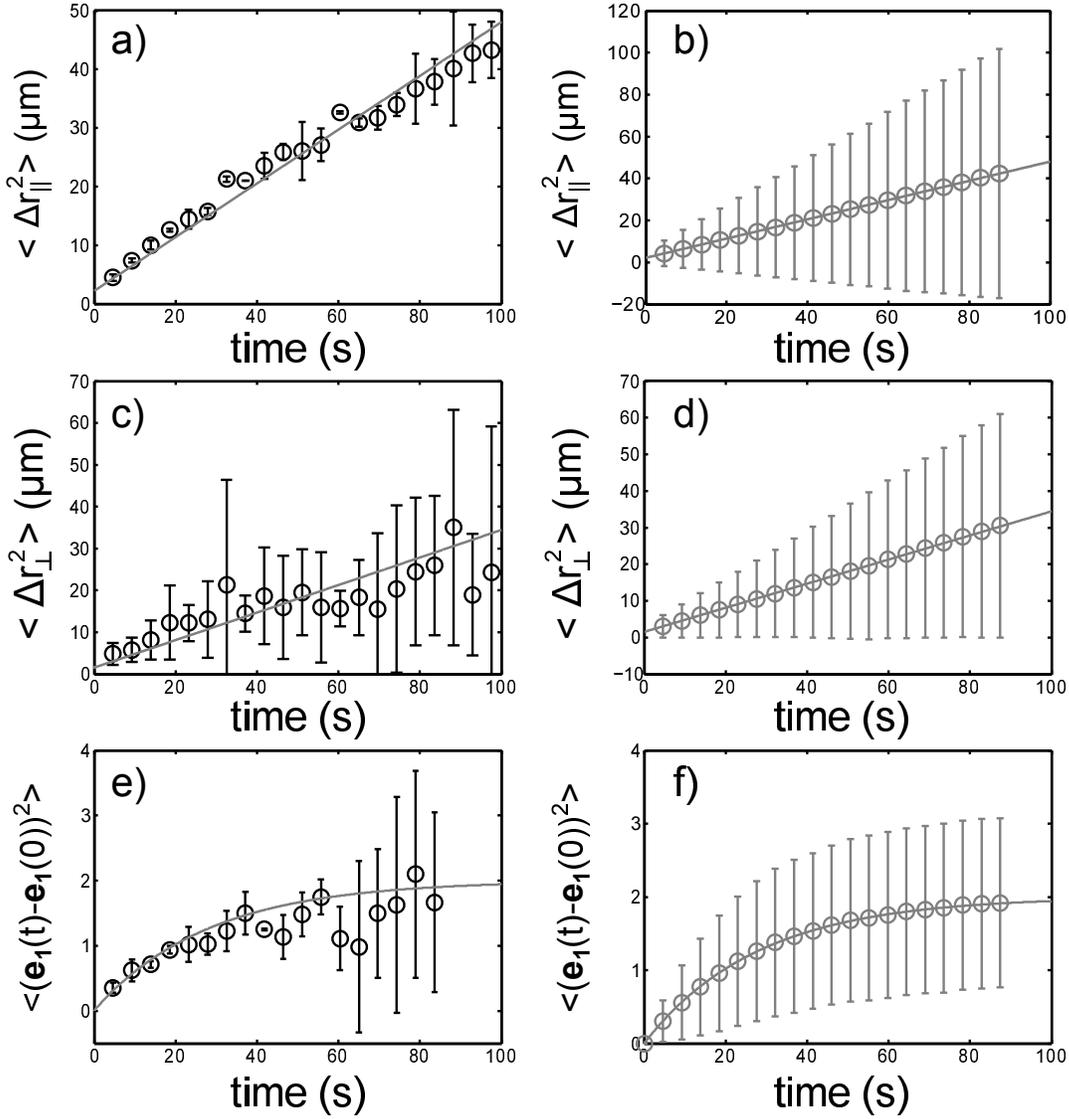}
\caption{The mean-squared displacement parallel to the major axis, $\left\langle\Delta r_{\Vert}^{2} \right\rangle$, (a-b), perpendicular to the major axis, $\left\langle\Delta r_{\perp}^{2} \right\rangle$, (c-d), and the rotational diffusion against the helix's major axis, $\left\langle(\mathbf{e_{1}}(t) -\mathbf{e_{1}}(0))^{2}\right\rangle$, (e-f), of a diffusing LI cell. The black circles (left column) are the experimental data produced by Butenko \textit{et al.}, the grey circles are the data points created through the numerical simulation of the experiment and the solid grey lines are the `optimal' diffusion lines found from the least squared regression of the numerical data using the same sampling method as in the experiments.}
\label{experiment_helix}
\end{center}
\end{figure}

The trajectories from the Brownian Dynamics computations  were then sampled every 4.6 seconds and processed similarly to the  experiments, as detailed in Sec.~\ref{sec:es}. This provides an estimated value for $\left\langle(\mathbf{e_1}(t) -\mathbf{e_1}(0))^{2}\right\rangle$, $\left\langle\Delta r_{\Vert}^{2} \right\rangle$ and $\left\langle\Delta r_{\perp}^{2} \right\rangle$ every 4.6 seconds. This data is plotted in the right column of figure~\ref{experiment_helix} (grey circles) with the standard deviation of the data set added as error bars. 
Assuming that we have $\left\langle\Delta r_{\Vert}^{2} \right\rangle = 2 D_{\Vert} t + c_{\Vert}$, $\left\langle\Delta r_{\perp}^{2} \right\rangle = 2 D_{\perp} t + c_{\perp}$, and $\left\langle(\mathbf{e_1}(t) -\mathbf{e_1}(0))^{2}\right\rangle = 2[1-\exp(-2D_{\theta}t)]$ and performing the least squared regression on the numerical data, the apparent experimental `diffusion coefficients' for LI are predicted to be
\begin{eqnarray}
D_{\Vert} &=& 0.230~\mu \mbox{m}^{2} s^{-1}, \\ 
c_{\Vert} &=&  2.16~\mu\mbox{m}^{2},\\ 
D_{\perp} &=& 0.164~\mu \mbox{m}^{2} s^{-1}, \\ 
c_{\perp} &=& 1.52 ~\mu\mbox{m}^{2},\\ 
D_{\theta} &=& 0.0178~\mbox{rad}^{2} s^{-1}, 
\end{eqnarray}
 We then compare on the left column of figure~\ref{experiment_helix}  the experimental data (black circles) with the least squared regression lines predicted from our simulations (grey lines). Note the excellent agreement between experiments and theory in Figure 7.
Remarkably, this agreement, which is the most important result of the present work, 
is achieved with no adjustable parameters.
Our model captures therefore accurately  all the relevant physics.

The error bars in figure~\ref{experiment_helix} are the standard deviation on the functions for both the experiment and the numerical simulations. Specifically, they are the square root of the variances of $(\mathbf{e_1}(t) -\mathbf{e_1}(0))^{2}$, $\Delta r_{\Vert}^{2}$, and $\Delta r_{\perp}^{2}$.  No additional component was added to the error bars to indicate error in the numerical calculation. Roughly speaking the variance on the mean squared displacement should be proportional to $t^{2}$ so that the standard deviation should increase linearly in $t$. The continuous growing spreads in figures~\ref{experiment_helix}b  and \ref{experiment_helix}d look  linear, supporting the idea that the error bars are the actual standard deviation and not a representation of the numerical error. 

The error bars on the experimental data points are noted to be much smaller than those on the numerical data. We believe that this is due to sampling limitations within the experiment where only a relatively small number of particles would have been tracked while the numerics has effectively tracked 10,000 particles. This effect could have been compounded by the variation in shape between different LI in the experiment. To handle the shape variation a larger number of particles must be tracked to obtain the full statistical ensemble.

We note that our model does not include the spiral or hook ends which is needed for LI mobility \cite{Nakamura2014}. In their experiment, Butenko \textit{et al.} typically chose to track specific LI cells without hooks. Our results further suggest that the spiral ends do not significantly affect the passive diffusion of LI. 

Finally, the data in figure~7 is replotted in  figure~8 where, this time, the mean square displacements have been divided by time to display explicitly the value of the diffusion constants. In addition we have also added the diffusion of \eref{dpara}, \eref{dperp}, and \eref{drot} to the figure (dashed lines).  The rotational diffusion is not replotted as the sampling has no effect. However the diffusion parallel and perpendicular do show a difference. Both dashed lines tend to underestimate the experimental results at early times. This is due to the experimental sampling creating an effective non-zero intercept. At later times, both the unsampled diffusion behaviours (dashed lines) seem to match the experimental behaviour well and lie very close to the sampled behaviour at these latter times. The experimental diffusion rate parallel does tend to be slower than the predicted rates however. The slower parallel diffusion rate and the need for a non-zero intercept are both results of the sampling and processing of the experimental data. This confirms that the simulation of the sampling is necessary to match the experimental behaviour.

\begin{figure}
\begin{center}
\includegraphics[width=0.9\textwidth]{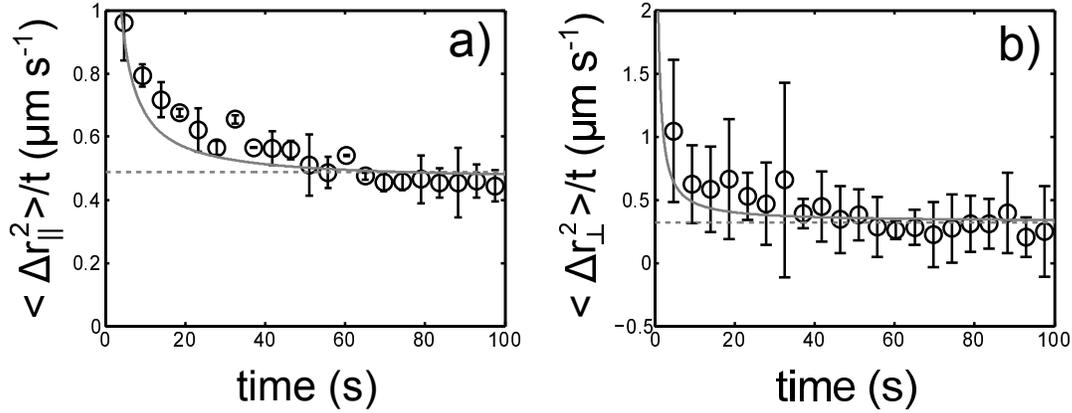}
\caption{The mean-squared displacement parallel to the major axis divided by time, $\left\langle\Delta r_{\Vert}^{2} \right\rangle / t$, (a), and perpendicular to the major axis divided by time, $\left\langle\Delta r_{\perp}^{2} \right\rangle / t$, (b) of a diffusing LI cell. The black circles are the experimental data produced by Butenko \textit{et al.}, the solid grey lines are the `optimal' diffusion lines found from the least squared regression of the numerical data using the same sampling method as in the experiments and the dashed grey lines are the diffusion behaviour without the experimental sampling.}
\label{experiment_against2}
\end{center}
\end{figure}

\subsection{Approximation by a prolate spheroid} \label{sec:ps}

\begin{figure}
\begin{center}
\includegraphics[width=0.9\textwidth]{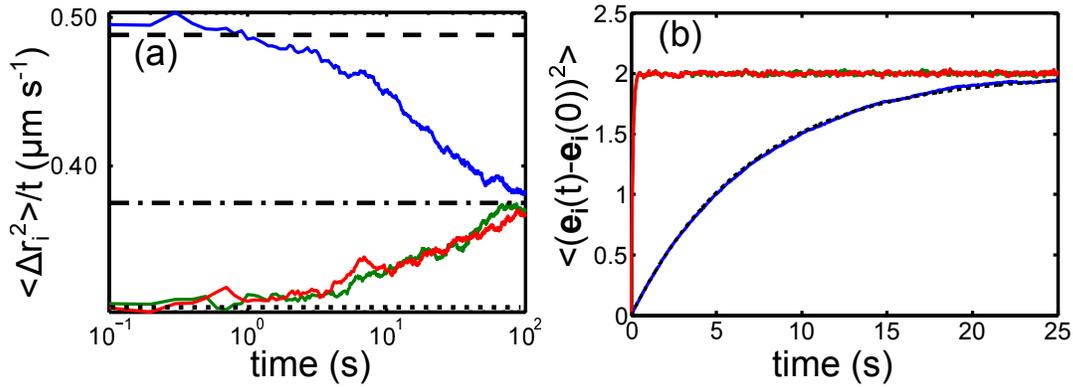}
\caption{(Colour online) Directional mean-squared displacement divided by time (a) and $\left\langle(\mathbf{e_{i}}(t) -\mathbf{e_{i}}(0))^{2}\right\rangle$ (b) for  our model of LI cell in the laboratory frame. The cell's major axis, $\mathbf{e_1}$, is initially aligned with (1,0,0) and  we have $k_{b} T /\mu = 3.70~\mu\mbox{m}^{3} s^{-1}$. 
 (a): the solid blue line represents the mean squared displacement in (1,0,0), the solid green in (0,1,0), the solid red in (0,0,1), while the black lines are the behaviors  around the major axis (dashed), minor axis (dotted) and the long-time diffusion constant (dot-dashed); 
(b) the black dashed line is the least squared regression line for how $\mathbf{e_1}$ should change, the solid blue line is the calculated change in $\mathbf{e_1}$, while the solid green and red lines show the dynamics of  $\mathbf{e_2}$ and $\mathbf{e_3}$ respectively.}
\label{helixexact}
\end{center}
\end{figure}

The dominance of the diagonal terms in \eref{RM} suggests that the LI helix is hydrodynamically similar to a prolate spheroid. This similarity is also seen in the laboratory frame diffusion of LI where the dynamics of a prolate spheroid shown in figure~\ref{prolatetest} is replotted in figure~\ref{helixexact} for the case of the LI helix.  The results are close to those for a prolate spheroid because  the helix radius, $r_{h}$, helix pitch, $P$, and body radius, $r_{b}$ are all much smaller than the helix length, $L$. 
As a result, we obtain very little velocity variation between adjacent loops in the helix thereby `blurring' the helical shape from the background flow. The prolate spheroid that best replicates the resistance matrix in \eref{RM}  has a semi-major and semi-minor axes of 4.60~$\mu$m and 0.108~$\mu$m, respectively. This is obtained   by minimizing the squared difference between the diagonal terms in \eref{RM} and the exact values of prolate spheroid \cite{Chwang2006} for the semi-major and semi-minor axis lengths of the prolate spheroid. The results were very close with a residual squared error of 0.698, two order of magnitude smaller than most of the diagonal terms. The close agreement indicates that it may be possible to model such a helix as an appropriately shaped prolate spheroid. The calculation further suggests that the prolate spheroid should have a semi-major axis of roughly half the body length, $L/2$, and a minor axis of about three times the body thickness, $3 r_{b}$. In this specific case $r_{b} \approx r_{h}$ and so $3 r_{b}$ is roughly the average of the diameter of the circle the helix body makes perpendicular to its major axis, $ 2 r_{h} + 2 r_{b} \approx 4 r_{b} $, and the diameter of the body itself, $ 2 r_{b} $.

\subsection{Comparisons with other models} \label{sec:om}

 The value of the diffusion coefficient parallel to the major axis, \eref{dpara}, is significantly larger than the predicted value of 0.1337 $\mu \mbox{m}^{2} s^{-1}$ by the old helix model by Hoshikawa \cite{Hoshikawa1979,Hoshikawa1976}. Hoshikawa's model treated the body as a series of spheres with radius $r_{b}$ which interact hydrodynamically through the stokeslet flow of each sphere's individual movement \cite{Hoshikawa1976}. The model assumes that there are many spheres in one helix pitch so that the force distribution can be approximated by a continuous distribution, and ignores any end effects. For a tightly coiled helix, like that of LI cells, there are around five spheres in one helix pitch breaking the many-spheres assumption. Furthermore, and as expected, by only interacting the spheres through stokeslet interactions the local behaviour of the helix is incorrectly represented. To estimate the significance of these local contributions, the resistance matrix of the LI helix was computed using  
 resistive-force-theory \cite{GRAY1955,Lighthill1987}, an approximation of slender body theory which  considers only the local contributions to the motion of a slender body. For the shape described by \eref{centreline}, the resistance matrix given by  resistive force theory is
\begin{eqnarray} \label{RMresistive}
\fl \mathbf{R}_{\tilde{F} \tilde{U}}^{RF}/\mu =  \left(
\begin{array}{c c c c c c}
 10.78 &  -0.031 &   0.026 &   0.39 &   0.18  & -0.14\\
   -0.031 &  11.04 &   0.020 &  -0.0045 &  -0.19  & -0.019\\
    0.026  &  0.020 &  11.05  &  0.0036  & -0.019  & -0.20\\
    0.39   &-0.0045  &  0.0036   & 0.057  &  0.026  & -0.020\\
    0.18   &-0.19   &-0.019    &0.026   &76.33  & -0.43\\
   -0.14   &-0.019  &-0.20   &-0.020   &-0.43  & 76.14
\end{array}\right),
\end{eqnarray}
where, as in  \eref{RM}, all lengths are in $\mu$m. Comparing  \eref{RMresistive} with  \eref{RM} one sees that about two thirds of the resistance coefficient for force parallel to the major axis due to motion in the same direction is coming from local contributions. The inability of the old helical model to represent the local contribution accurately would thus, expectedly,  create a large deviation. The values in  \eref{RMresistive} also display three other important differences when compared with those in  \eref{RM}: the off-diagonal terms are typically larger, the resistance coefficient for force perpendicular to the major axis due to motion in the same direction is half the size of the slender body prediction, and the resistance coefficient for torque perpendicular to the major axis from rotation in the same direction is much smaller in the resistive force case. These differences are due to a combination of  end effects and long-range interactions which are properly included in  slender body theory. In particular,  long range interactions will be the leading cause of the prolate spheroid-like behavior from  \eref{RM}. Therefore the addition of long-range interactions would reduce the size of the coupling terms. Similarly, 
long-range interactions would increase the resistance coefficients for both the perpendicular motion and rotation as it reduces flow through the adjacent loops of the helix. End effects could be very important for the rotation perpendicular  to the cell axis because the ends  produce higher torque then central points and so could seriously affect the results. In order to  accurately describe and predict the three-dimensional diffusion of LI cells,  both  long-range hydrodynamics interactions and   end effects  are thus  important.

\section{Conclusions}

In this paper the behaviour of a tightly wound helix undergoing Brownian motion was numerically investigated. The dimensions of the helix were chosen to closely reflect the shape of \textit{Leptospira interrogans} cells. The resistance matrix of the helix was determined numerically using a validated implementation of  slender body theory, and was then exploited in Brownian Dynamics to describe the thermal diffusion. The statistical results  are in excellent quantitative agreement with the experimental results of Butenko \textit{et al.} \cite{Butenko2012} (figures~\ref{experiment_helix} and \ref{experiment_against2}), showing that the model accurately describes the needed physics with no adjustable parameters. The diffusion of the tightly wound helix is seen to closely reflect the diffusion of a prolate spheroid. This similarity can be seen in the helix's resistance matrix whose diagonal terms are  very similar to those of a prolate spheroid with a semi-major and semi-minor axis of 4.60~$\mu$m and 0.108~$\mu$m.  The misrepresentation of the local and long-range contributions was also found to be the probable failure of the old helix model.

The method employed for the diffusion of \textit{Leptospira} could be used to investigate the diffusion of any arbitrarily-thin rigid body. Substituting the centerline description   into the slender body theory equations will give a resistance matrix which, when written in the centre of mobility frame, can be used to determine the diffusion from the Langevin equations. Similarly, the method to simulate the experimental sampling can be extended to compensate for arbitrary experimental sampling times and diffusion calculation procedures. The diffusion model could also be adapted to look at the diffusion of actively swimming particles of realistic shapes, which  will be the subject of future work.

\ack
The authors  thank Alexander V. Butenko and Eli Sloutskin for sharing their results and helpful discussion about their experimental procedures. This research was funded in part by the European Union (CIG grant to E.L.).

\section*{References}

\bibliographystyle{unsrt}
\bibliography{library}

\end{document}